\documentclass{article}
\usepackage[fleqn]{amsmath}
\usepackage{amssymb}
\usepackage{epsfig}
\usepackage{epsf}

\newtheorem{definition}{Definition}

\newtheorem{condition}{Condition}

\title{Designing a Bayesian Network for Preventive Maintenance from Expert Opinions in a Rapid and Reliable Way}
\author{G. Celeux\thanks{INRIA Futurs 91405 Orsay cedex FRANCE}, F. Corset\thanks{LabSAD, EA 3698, UPMF, 1251 Avenue centrale BP47 38040 Grenoble cedex 9,FRANCE}, A. Lannoy\thanks{EDF R\&D D\'{e}partement MRI 6, quai Watier, 78401 Chatou cedex, FRANCE} and B.Ricard\footnotemark[3]\\
}
\date{}
\begin{document}
\maketitle

\begin{abstract}
In this study, a Bayesian Network (BN) is considered to represent a nuclear plant
mechanical system degradation. It describes a causal representation
of the phenomena involved in the degradation process. 
Inference from such a BN needs to specify
a great number of marginal and conditional probabilities. As, in the present context,
information is based essentially on expert knowledge, this task
becomes very complex and rapidly impossible. We present a solution
which consists of considering the BN as a log-linear model on
which simplification constraints are assumed. This approach
results in a considerable decrease in the number of probabilities
to be given by experts. In addition, we give some simple rules to choose the most reliable probabilities. 
We show that making use of those rules allows to check the consistency of the derived probabilities. 
Moreover, we propose a feedback procedure to eliminate inconsistent probabilities.
Finally, the derived probabilities that we propose to solve the equations
involved in a realistic Bayesian network are expected to be reliable. 
The resulting methodology to design a significant and powerful BN is applied to a
reactor coolant sub-component in EDF Nuclear plants in an illustrative purpose.
\end{abstract}


\section{Introduction}
Preventive maintenance is considered in a lot of industries because costs due to failures and repairs could be very
important. Moreover, since system safety is an important goal for
industries, the knowledge of the degradation processes has became
essential. 
Preventive maintenance of a system is taking into account expert knowledge, feedback observations and degradations in order 
\begin{itemize}
\item to model the system lifetime and to quantify the degradation
or failure probability, 
\item to detect important variables
involved in the degradation process and to design maintenance tasks
in order to differ or eliminate ageing, 
\item to quantify the effect of maintenance actions on the system behavior, \item to
propose diagnosis and decision help, \item to propose data
mining and sensibility analysis.
\end{itemize}
Bayesian Networks, abbreviated BNs in the following, could be thought of as
useful to help engineers to fulfill those purposes of preventive maintenance.
BNs provide an efficient way to represent the degradation process of an industrial system or machine.
Bayesian Networks are some
specific graphical models 
introduced by Pearl \cite{PEA88}, and Lauritzen and Spiegelhalter
\cite{LAU88}. Graphical Models introduce a relation between graph 
and probability theories. The random variables of a probabilistic model are
described with the vertices of a graph, where edges describe their dependencies measured with conditional
probabilities. A great interest of BNs is to provide an efficient tool for
modelling in a simple and readable way the most probable links between events of different nature
(expert opinion, feedback experience, \ldots) using conditional independence between random variables. 
BNs which can also be regarded as a way to introduce randomness in influence diagrams can be expected to
be useful for preventive maintenance in the future. The articles of \cite{SAC00} and \cite{VAT97}
are recent examples of such maintenance modelling with influence diagrams.\\

Bayesian Networks, by describing the main conditional probabilities between
variables, allow to compute easily the joint probability distribution of all the variables
involved in a complex process.
However, in order to obtain this joint probability distribution,
the number of required probabilities increases exponentially with the number
of variables in the model. In the last decade, authors developed
algorithms  in order to facilitate calculations in graphical
models. We can cite the well-known "junction tree" algorithm (see
Lauritzen \cite{LAU88} (1988), Cowell \cite{COW99b} (1999) for instance). Thus
many efficient algorithms to deal with
computations on more and more complex graphical models are available. These
algorithms are included in many software programs like, for
instance, the Denmark software Hugin expert \cite{HUGIN} and Netica \cite{NETICA} of
Norsys Software Corp. If it is essential to develop efficient
algorithms to take full advantage of the knowledge contained in
graphical models, we must keep in mind that if the knowledge base
is poor or badly managed the best algorithm becomes useless (see
the introduction section in Cowell et al. \cite{COW99a}). 
The aim of the present paper is to deal with the practical difficulties encountered when designing a
BN for a real maintenance modelling problem. All the ideas we present are motivated and illustrated 
with the modelling of a nuclear mechanical system degradation with a Bayesian Network. 

First, we address the problem of deriving reliable information from experts
in a graphical model setting and propose a method to obtain honest
probability values in a simple and possibly interactive way from
experts knowledge. 
Then we deal with a second difficulty. In order to compute the joint probability of the variable, 
the classical BN approach consists of asking for the marginal probabilities 
of the entry variables and the conditional
probabilities of the other variables knowing all the possible
combinations between their parents. 
Owing to the great number of parameters a BN can invovle in practice, it can become a formidable task, and we propose to approximate the graphical model with an
unsaturated log-linear model (see for instance \cite{WHI90}, chapter 7).
This procedure leads to consider that some variables are conditionally
independent. In the maintenance modelling context, 
the experts are asked to provide the marginal probabilities of all the
variables and not only of the parent variables as in the classical
approach.
We choose this method because generally experts can more easily provide such simpler
marginal probabilities than conditional probabilities. This strategy leads to a system with more equations
than unknown quantities. Thus, it allows us
to assess the consistency of the required probabilities. According to Bayes theorem, the properly weighted summation of conditional
probabilities of a vertex $N$, knowing a parent vertex,
is equal to its marginal probability. After checking the consistency of the
given probability values, a feedback procedure is proposed when
some inconsistency is encountered, and the experts can choose the
reliable given probabilities from their own viewpoint. Thus, the
probabilities finally selected to solve the system of $n$ equations
with $n$ unknown quantities are expected to be the most reliable ones.\\ 

This paper is devoted to present our heuristic strategy
to build practically a graphical model in a realistic and relevant way. It is organized as follows. 
In Section 2, Bayesian Networks models are introduced and a parallel
between graphical models and log-linear models which is helpful to reduce the BNs complexity is presented. 
Section 3 is devoted to the presentation of our methodology to get information
from experts in a simple and reliable way. We indicate
possible drawbacks of an over restrictive strategy and present
ways to remedy these drawbacks. An illustrative application concerning nuclear
mechanical system degradation is presented in Section 4 and a short
conclusion section ends this paper.

\section{Bayesian Networks}
Bayesian Networks are powerful graphical models to describe
conditional independence and analyze probable causal influence
between random variables. In our study, variables are all discrete
and most of them are binary. These random variables are represented by the
vertices of the graph, and the probable influence between two
random variables is represented by an edge between the
corresponding vertices. We now give some definitions.

\begin{definition}
A directed graph is a couple $\mathcal{G}=(V,E)$, where
$V=(X_1,\ldots,X_n)$ denotes the vertices of the graph and
$E=(e_1,\ldots,e_m)$ denotes a part of Cartesian product $V\times
V$, where $e_i$ is called the edges of the graph.
\end{definition}
If $(X_i,X_j)$ lies in $E$, then this element is called an edge.
It is
denoted $X_i \longrightarrow X_j$, $X_i$ is called the source and $X_j$
the target of the edge. For directed graph, the parents and the
children of the vertices are defined as follows:
\begin{definition}
If a directed edge has source $X_i$ and target $X_j$, then $X_i$ is called
the parent of $X_j$ and $X_j$ is called the son or child of $X_i$. The set of the parents of $X_j$ is
denoted $pa(X_j)$and the
set of children of $X_i$ is denoted $ch(X_i)$.
\end{definition}

In a directed graph, the oriented paths are defined as follows:
\begin{definition}
An oriented path is a set of distinct vertices $X_i,\ldots,X_j$
such that $(X_{k-1},X_k)$ is an edge for all $k=i+1,\ldots,j$.
This path is denoted $X_i \longmapsto X_j$. A cycle is a path such
that $X_i=X_j$.
\end{definition}
Directed graph without cycle are called Directed Acyclic Graphs
(DAG). We can now define a Bayesian Network.

\begin{definition}
A Bayesian Network is
\begin{itemize}
\item a set of variables $V$, defining the vertices, and a set of edges between variables
$E$, \item each variable has a finite number of exclusive states,
\item variables and edges define a directed acyclic graph,
denoted $G=(V,E)$, \item for each variable $Y$ with its parents
$X_1,\ldots,X_n$, is associated a conditional probability
$P(Y|X_1,\ldots,X_n)$. When a variable has no parent, the last
quantity becomes a marginal probability $P(Y)$.
\end{itemize}
\end{definition}

The denomination "Bayesian Networks" comes from the well-known
Bayes theorem.
In a BN, the joint probability can be written as
follows (recursive factorization):
\begin{equation}
\label{RECFAC} P(X_1,\ldots,X_n) = \prod_{i=1}^n P(X_i|pa(X_i)),
\end{equation}
where $pa(X_i)$ is the set of parents of vertex $X_i$.\\ 

A first problem when designing a BN is to define sensible vertices between the variables involved in the study. 
This task is made more difficult when numerous variables are available. In our application, in order to deal with this difficulty, 
we follow a strategy used in H\o{jsgaard} \cite{HOJ96} to build the structure of the BN. 
This strategy consists essentially of grouping variables playing analogous roles
(see \cite{COR03}).\\

In order to draw useful informations from the resulting BN, the knowledge of the joint probability is required. 
Due to the lack of data, we essentially use experts opinion and we use data when it is possible. However, even 
with binary variables in Equation (\ref{RECFAC}), it is important to notice that if a vertex has a lot of parent 
vertices (more than two parents for instance), there are a lot of conditional probabilities to be estimated. In this situation, 
additional help of experts is needed. But, experts can only give simple probabilities, and it is out of matter to 
ask them a conditional probability knowing a great number of possible combinations. To overcome this second difficulty, 
the BN is seen as an unsaturated log-linear model. It is well-known that a BN is always equivalent to an unsaturated log-linear model,
where the conditional independences are derived from the moral graph of the BN.
Notice that conversely  a log-linear model needs that Condition \ref{chris} stated below is verified
to be representable with a BN.
Log-Linear models allow to analyze the relationships between qualitative variables in a contingency table 
(See Goodman \cite{GOO70}, Bishop et al. \cite{BIS75} and Christensen \cite{CHR90}). These models are 
flexible, currently available in many statistical softwares and have a powerful interpretation in 
conditional independence terms. With this representation in mind and in order to reduce the number of required 
conditional probabilities (i.e. the complexity of the BN), some new conditional independences are added.\\

\begin{figure}[h]
\begin{center}
\includegraphics[scale=1]{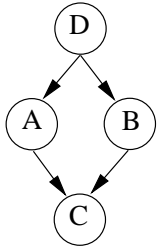}
\caption[Bayesian Network with four variables]{\label{graphABCD}{Bayesian Network with four variables.}}
\end{center}
\end{figure}

In an illustrative purpose, consider the BN depicted in Figure \ref{graphABCD}.

Figure \ref{graphABCD} exhibits the conditional independence of $A$ and
$B$ knowing $D$, and of $C$ and $D$ knowing $A$ and $B$. Now consider the associated unsaturated log-linear model:
\begin{eqnarray}
\label{SatLogLin}
\log(p(a,b,c,d)) & = & u+ u_A(a) + u_B(b) + u_C(c) + u_D(d) \nonumber\\
       & + & u_{AB}(a,b) + u_{AC}(a,c) + u_{BC}(b,c) + u_{ABC}(a,b,c)\nonumber\\
       & + & u_{AD}(a,d) + u_{BD}(b,d),
\end{eqnarray}
with the interaction terms $u$, called $u$-terms, being functions
of the cell probabilities of the four-way contingency table. We denote this model $[ABC][AD][BD]$. In this model, $C$ and $D$
are conditionally independent knowing $A$ and $B$. These two representations are strictly equivalent. A possible way to reduce 
the complexity of the model is to assume that $A$ and $B$ are conditionally independent knowing $C$.
It leads to delete the term $u_{ABC}$ 
in Equation (\ref{SatLogLin}). This new unsaturated log-linear model is making sense, but we are facing a problem because
it cannot be represented with a BN. This problem can easily be solved by using the following condition (see for instance
Christensen \cite{CHR90}).
\begin{condition} \label{chris}
A log-linear model can be represented by a Bayesian Network if,
whenever the model contains all two-order interaction terms
generated by a higher-order interaction terms, then the model
contains this higher-order interaction terms.
\end{condition}

With a more complex BN, for a vertex with a lot of parents, we think that it is impossible to manage all conditional dependences. Thus, to reduce the complexity of a causal BN, we propose to first assume that all $u$-terms, with order greater than two, are equal to zero. 
It is equivalent to consider {\em conditional independence between parent vertices knowing their son}.
As this assumption could be too restrictive, in a second stage, we ask the experts to add some higher order associations 
(three-way interaction terms):The ones they considered useful and reliable.
In practice, this additional complexity required by condition (\ref{chris}) to make the log-linear model representable with 
a BN does not appear to be too important because it is not judicious to model associations of order larger than 
three. It is important to note that if the expert adds an association, he has to quantify this association by giving the corresponding conditional probabilities.\\

After this complexity reduction step has been completed, we can design a strategy to extract information from data and experts.

\section{Information Extraction Strategy}
As seen from Equation (\ref{RECFAC}), the joint probability of a BN can be
written in a recursive factorization form. Thus, for all vertices,
it is necessary to evaluate the conditional probabilities of the
variables knowing their parents. Most of the probabilities are given by
expert opinions. The statistician must prepare a questionnaire for the
experts. At each probability to be given corresponds a question, written in simple words,
without technical stuff. Moreover, the interviews are completed expert by expert.
This organization allows to avoid correlations between
experts answers.\\

The main difficulty for a non-statistician expert is to apprehend
the conditional probability concept. For instance, let us consider a BN with
three binary variables, $S$, $A$ et $C$, respectively smoker,
alcoholic, and cancer of throat. Experts are able to give the
probability that someone have a cancer of throat knowing that he
is a smoker, and the marginal probability to have this cancer. But,
the probability that someone has a cancer
knowing he is not a smoker is more difficult to evaluate. Most of
experts tend to give, for this last probability, the marginal
probability to have cancer, considering implicitly that to be a non-smoker is
not an information. In this example, experts must be able to
give four conditional probabilities, as the probability that
someone has a cancer of throat, knowing that he is a non-smoker
and a non-alcoholic. This combination is relatively simple for a
vertex with two parents, but it seems difficult to give all
conditional probabilities with a greater number of parents.\\

When preparing the questionnaire, it is important not to forget rare events. 
For instance, let us consider the case of a
very reliable component with a probability of default equal to
$10^{-6}$. Assume a Bayesian Network with two binary variables:
"default" (yes/no) and "state" (healthy/failure), where the
variable "state" is a probable consequence of the variable
"default". In order to compute the joint probability, it is
necessary to evaluate the marginal probability of ``default'' and
the two conditional probabilities of the ``state'' knowing the
``default'' variable. In this case, it is clearly preferable to query the conditional
probability of failure knowing the presence or the absence of default, because both probabilities are significantly different.\\

The evaluation of all required probabilities, especially 
conditional probabilities, becomes rapidly an impossible task for 
experts. Thus, it is of primary importance to give simple
rules in order to realize this preliminary step (see
\cite{CHA00}). Hereunder, we summarize the rules we have chosen according to the above mentioned considerations.\\

\noindent
{\bf Rule 1: Query all marginal probabilities}.
Considering that marginal probabilities are easier to be 
evaluated, these probabilities are asked even for non input
variables. In the classical procedure, only marginal probabilities
of input variables are required. However, evaluating all the marginal probabilities
does not allow to compute the joint
probability. Generally, some conditional probabilities are
needed, except if variables are all independent. In the binary
case, the number of conditional probabilities is $2^n$, where $n$ is
the number of parents. It is impossible to evaluate these
probabilities as soon the number of parents is greater than three or four.\\

\noindent
{\bf Rule 2: Query only the conditional
probabilities of first order}. For instance, let us consider a Bayesian
Network with three parent vertices $A$, $B$, $C$ and a son node $D$.
For inference, three marginal probabilities are
required ($p(A)$, $p(B)$, $p(C)$) and $p(D|ABC)$, corresponding to
eight conditional probabilities.  in a first approximation, our methodology consists of querying
the four marginal probabilities and the following conditional
probabilities: $p(D|A)$, $p(D|\bar{A})$, $p(D|B)$, $p(D|\bar{B})$, $p(D|C)$, $p(D|\bar{C})$.
These probabilities are usually the simplest ones to give for non-statistician
experts. Moreover, one can see that those probabilities involve a lot of
redundancies. Thus, this method allows to turn out the incoherences in
the expert allocation. Indeed, we have
\begin{equation}
p(D)= \sum_{A} p(D|A)p(A)=\sum_{B} p(D|B)p(B)=\sum_{C} p(D|C)p(C).
\end{equation}
In practice, these equations are never exactly verified. The statistician
must decide to eliminate some of them. In the example below, one
can choose to remove three conditional probabilities. Keeping in
mind that conditional probabilities knowing a "non-event" are
difficult to evaluate, thus statistician should remove
$p(D|\bar{A})$, $p(D|\bar{B})$ and $p(D|\bar{C})$. However, this
rule is not an absolute rule. Hereafter, we propose others rules to decide
which probabilities have to be removed.\\

\noindent
{\bf Rule 3: The most relevant probabilities come from databases}.
We consider that the feedback experience is more reliable
than the expert opinion, especially for conditional
probabilities. Moreover, when feedback experience is available, experts
based generally their opinion on this feedback experience. In such a
situation, some marginal probabilities can be derived wihout too much difficulty, but unfortunately for
some variables, with a lot of parent vertices, estimating a
conditional probability requires a lot of data.\\

\noindent
{\bf Rule 4: Favor marginal probabilities provided by the experts}. 
This rule arises from the fact that those probabilities are easier 
to evaluate, in particular for a non-statistician expert.  Thus, it remains to choose one of the conditional probabilities per parent vertex.

\begin{figure}[h]
\begin{center}
\includegraphics[scale=1]{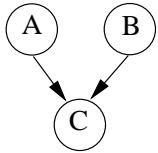}
\caption[BN with two parent vertices]{\label{graph_2p}{BN with two parent vertices.}}
\end{center}
\end{figure}

For instance, let us consider a Bayesian Network with three vertices as in Figure \ref{graph_2p}.
A possible strategy consists of requiring to which probabilities the expert is more confident. However, in most cases, the choice is very 
restricted. Indeed, let us suppose that for the BN in Figure \ref{graph_2p} where $A$ is a three-level variable, the expert give probabilities 
displayed in Table \ref{ExemplePro}.
\begin{table}[h]
\begin{center}
\begin{tabular}{|c|c|}
\hline
$P(C)=0.25$ & \\
\hline
$P(C|A=0)=0.05$ & $P(A=0)=0.33$ \\
\hline
$P(C|A=1)=0.25$ & $P(A=1)=0.66$ \\
\hline
$P(C|A=2)=0.30$ & $P(A=2)=0.01$ \\
\hline
\end{tabular}
\caption[Example of probabilities given by an expert]{\label{ExemplePro}{Example of probabilities given by expert.}}
\end{center}
\end{table}
From Table \ref{ExemplePro}, it is possible to compute the probability of $C$:
\begin{equation*}
P_{calc}(C) = 0.05*0.33+0.25*0.66+0.30*0.01=0.183,
\end{equation*}
which is different of the probability given by the expert. Assume now that, in order to remove this gap, the expert decides to evaluate 
the last conditional probability from an algebraic calculus. It leads to the new conditional probability
\begin{equation*}
P_{calc}(C|A=2) = \frac{0.25-0.05*0.33-0.25*.66}{0.01} = 6.85 !
\end{equation*}
This anomaly comes from the fact that the expert changes the conditional probability which has the lowest weight, namely
the weight corresponding to the marginal probability of the parent
vertice. Moreover, the highest conditional probability $P(C|A=1)$ with weight of $P(A=1)=0.66$ is equal to the marginal probability and the other conditional 
probability $P(C|A=0)$ with a significant weight $P(A=0)=0.33$ is rather low. Thus, to remove the difference between the two 
marginal probabilities (between $0.183$ and $0.25$), the conditional probability is strongly increased and exceeds one \ldots\\

\noindent
{\bf Rule 5: It is more convenient to change conditional probabilities with large weights, if those probabilities are significantly
different of the corresponding marginal probability}.\\

In the previous example, the application of rule 5 gives 
\begin{equation*}
P_{calc}(C|A=1) = \frac{0.25-0.05*0.33-0.30*0.01}{0.66} = 0.3492
\end{equation*}
which replace the previous value of $0.25$. Note here that the ranking of the three conditional probabilities has changed. 
To preserve the ranking and the ratio between the conditional probabilities, we can change all conditional probabilities 
by keeping the ratio between the given probabilities.\\

Assume now that an expert gives the probabilities displayed in  Table \ref{ExemplePro2}.
\begin{table}[h]
\begin{center}
\begin{tabular}{|c|c|}
\hline
$P(C)=0.05$ & \\
\hline
$P(C|A=0)=0.01$ & $P(A=0)=0.33$ \\
\hline
$P(C|A=1)=0.03$ & $P(A=1)=0.66$ \\
\hline
$P(C|A=2)=0.05$ & $P(A=2)=0.01$ \\
\hline
$P(C|B=0)=0.10$ & $P(B=0)=0.10$ \\
\hline
$P(C|B=1)=0.03$ & $P(B=1)=0.90$ \\
\hline
\end{tabular}
\caption{\label{ExemplePro2}{Another example of probabilitiesgiven by
    an expert.}}
\end{center}
\end{table}
The marginal probability can be seen as a convex combination of the conditional probabilities, where the weights are the marginal probabilities 
of the vertices parents. Thus, in the fourth line of Table \ref{ExemplePro2}, the conditional probability is equal to the marginal probability, 
$P(C|A=2)=P(C)$. Moreover, two other conditional probabilities ($P(C|A=0)$ and $P(C|A=1$) are strictly lower than the marginal probability. 
Thus, the weight of the conditional probability $P(C|A=2)$, namely $P(A=2)$ must be equal to one. In this particular case, we propose to 
change the marginal probability of $C$.
The probability $P(C)$ can be computed
from the two vertices parents
\begin{eqnarray*}
P^{A}_{calc}(C) & = & 0.029,\\
P^{B}_{calc}(C) & = & 0.037.
\end{eqnarray*}
The first probability is not a convex combination of the conditional probabilities of $C$ knowing $B$, i.e. $0.0029\notin [0.03;0.10]$. 
Thus, the second computed probability is to be preferred, and the same rules as previously defined are observed in order to change one 
of the conditional probabilities of $C$ knowing $A$.\\

Finally, it is possible to keep ratio between conditional probabilities by solving the following linear program 
\begin{equation*}
\min_{0\leq x\leq 1}|P(C)-\sum_{i=1}^3 k_i x P(A=i)|
\end{equation*}
where $k_i$ represents the ratio between the conditional probabilities given by experts.\\

In the last example, if no assumption is made on conditional independence, experts must give the conditional probability of $C$ knowing $A$ 
and $B$. Thus, we apply the same rules for these conditional probabilities, keeping in mind that we asked first the conditional probabilities 
of $C$ knowing $A$ and the conditional probabilities of $C$ knowing $B$. If the number of parents vertices is $n$ and if associations
with order greater than three are considered, the conditional probabilities can be computed by induction.

\section{Application}
The above described methodology has been applied to build a BN for a sub-component of a Reactor Coolant
Pumps, observed on the French Nuclear Plants. In this context, the choice of the system and the experts is crucial. 
First, the study must concern a system which the experts considered
significant for the safety of the whole industrial entity (here a nuclear plant). 
Moreover, the system must be well-known by the experts and the availability of the experts to build the structure of the 
graph must be a significant criterion, because this kind of project could be long and demanding and need many meetings. 
It is important to underline that the statistician have to play a role of mediator. He has to clearly define the goal of the project 
and describe the various powerful possibilities of this kind of model.\\

In the present case, the main goal of the study was to model the system degradation with a BN 
in order to define the most appropriate maintenance tasks and to evaluate the possible effects of this maintenance tasks.
In what follows we sketch the important points of this study according to the above described modelling and indicate
the practical interest of this study. \\

\begin{figure}[ht]
\begin{center}
\includegraphics[scale=1]{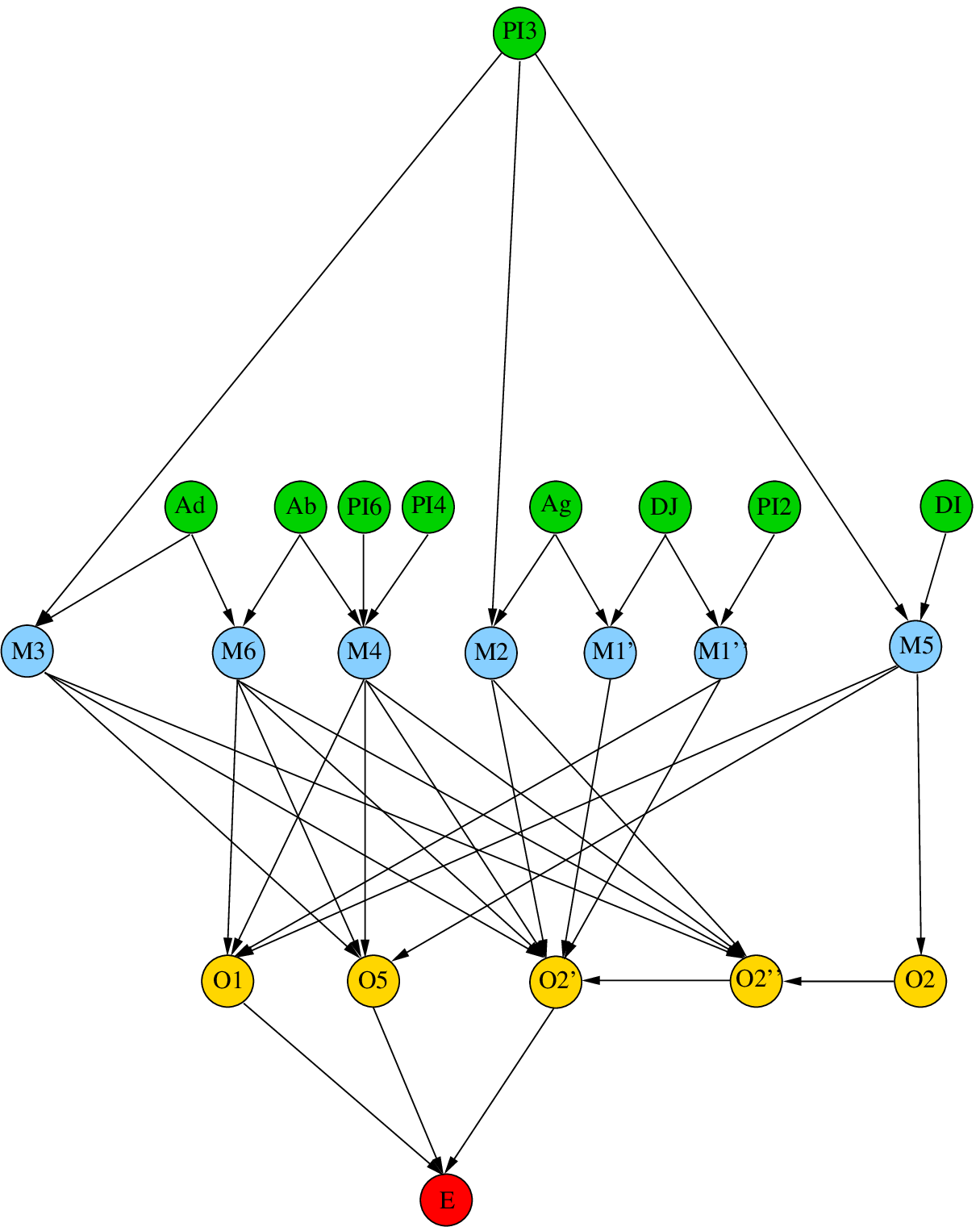}
\caption[Bayesian Network of a system degradation
process]{\label{joint1}{Bayesian Network of a system degradation
process.}}
\end{center}
\end{figure}

In Figure \ref{joint1}, we present a BN, built from experts opinions of the research and development
division of EDF according to the strategy described in \cite{HOJ96}. This 
Bayesian Network contains 22 discrete variables, 17 variables are binary and 
the other five ones have three modalities. According to the strategy of H\o{jsgaard} \cite{HOJ96}, variables have been divided in four sets:
environmental variables $\mathcal{A}=\{PI3,Ad,Ab,PI6,PI4,Ag,DJ,PI2,DI\}$, degradation variables 
$\mathcal{M} = \{M1^{'},\- M1^{''}, M2, M3, M4, M5, M6\}$, observation variables $\mathcal{O}=\{O1,O2,O2^{'},O2^{''},O5\}$ and
finally the variable of interest: the state of the system ($E$).\\

From the Bayesian Network presented in Figure \ref{joint1}, the likelihood of the model is given by
\begin{eqnarray}
\label{loijointe}
p(U) & = & p(Ab) p(Ad) p(Ag) p(PI2) p(PI3) p(PI4) p(PI6) p(DI)
p(DJ) \nonumber\\
 & \times & p(M1^{'}| Ag, DJ) p(M1^{''}| DJ, PI2) p(M2|Ag, PI3)
 P(M3|Ad, PI3)\nonumber\\
 & \times & p(M4|Ab, PI4, PI6)p(M5| DI, PI3) p(M6|Ad, Ab)\nonumber\\
 & \times & p(O1|M1^{''}, M4, M5, M6) p(O5|M3, M4, M5, M6) p(O2|M5)\nonumber\\
 & \times & p(O2^{''}|M2, M3, M4, M6, O2) \nonumber\\
 & \times & p(O2^{'}|M1^{'}, M1^{''}, M2, M3, M4, M6,
 O2^{''}) p(E|O1, O5, O2^{'}).
\end{eqnarray}

Inference from this BN requires the estimation of 381 probabilities in Equation (\ref{joint1}).
Now, since few data from feedback experience are available, experts must give a lot of probabilities 
and especially a lot of conditional probabilities. For instance, for the node $O2^{'}$, the classical 
approach requires to evaluate 192 conditional probabilities, a formidable task to deal with. Thus, in order 
to make the inference step feasible, we regard the BN as an unsaturated log-linear model, where all association 
terms of order greater than two are constrained to be zero. As stated above, this point of view is equivalent 
to assume that all parent nodes (variables) are conditionally independent knowing their sons. To evaluate 
all the probabilities needed for the computation of the joint probability, 
we proceeded by induction from the environmental variables to the variable of interest.\\

For instance, $p(M6\mid Ab,\;Ad)$ had to be computed in Equation (\ref{joint1}). Assuming that $Ad$ and $Ab$ are conditionally independent knowing $M6$, $P(Ab)$, $P(Ad)$, $P(M6)$, $P(M6\mid Ab)$ and $P(M6\mid Ad)$ were needed to compute $P(M6\mid Ab,\;Ad)$.

\begin{eqnarray*}
P(M6\mid Ad,Ab) & = & \frac{P(Ad,Ab\mid M6)P(M6)}{P(Ab)P(Ad)} = \frac{P(Ad\mid M6)P(Ab\mid M6)P(M6)}{P(Ab)P(Ad)}\nonumber\\
&  = & \frac{P(M6\mid Ab)P(M6\mid Ad)}{P(M6)}.
\end{eqnarray*}
The above formula is quite simple since $Ab$ and $Ad$ are conditionally independent knowing $M6$, but $P(M6)$ had to be provided. 

In the same manner, to compute $p(O1|M1^{''}, M4, M5, M6)$, by assuming that $M1^{''},M4,M5,M6$ are conditionally independent knowing $O1$, we get
\begin{eqnarray*}
P(O1\mid M1^{''},M4,M5,M6)  =  \frac{P(M1^{''},M4,M5,M6\mid O1)P(O1)}{P(M1^{''},M4,M5,M6)}\\
 =  \frac{P(M1^{''}\mid O1)P(M4\mid O1)P(M5\mid O1)P(M6\mid O1)P(O1)}{P(M1^{''},M4,M5,M6)},
\end{eqnarray*}
with 
\begin{equation*}
P(M1^{''},M4,M5,M6) = P(M1^{''})P(M5)\sum_{Ab}P(M4\mid Ab)P(M6\mid Ab)P(Ab),
\end{equation*}
where $M4$ and $M6$ are not independent, but are conditionally independent knowing $Ab$. Finally, $P(O1\mid M1^{''},M4,M5,M6)$ could be written 
\begin{equation*}
\begin{array}{l}
P(O1\mid M1^{''},M4,M5,M6)  =  \\
\displaystyle\frac{P(O1\mid M5)P(O1\mid M1^{''}) P(O1\mid M4) P(O1\mid M6)P(M4)P(M6)}{P(O1)^3 \sum_{Ab}P(M4\mid Ab)P(M6\mid Ab)P(Ab)}.
\end{array}
\end{equation*}

The last example concerns the variable of interest $E$ and the conditional probability $P(E\mid O1,O5,O2^{'})$:
\begin{eqnarray*}
P(E\mid O1,O5,O2^{'}) & = & \frac{P(O1,O5,O2^{'}\mid E)P(E)}{P(O1,O5,O2^{'})}= \\
 & = & \frac{P(O1\mid E)P(O5\mid E)P(O2^{'}\mid E)P(E)}{P(O1,O5,O2^{'})}\\
 & = & \frac{P(E\mid O1)P(E\mid O5)P(E\mid O2^{'})}{P(E)^2 P(O1,O5,O2^{'})}
\end{eqnarray*}
with
\begin{eqnarray*}
P(O1,O5,O2^{'}) & = & \sum_{\mathcal{M}} P(O1,O5,O2^{'}\mid \mathcal{M})P(\mathcal M)\\
 & = & \sum_{\mathcal{M}}P(O1\mid \mathcal{M})P(O5\mid \mathcal{M})P(O2^{'}\mid \mathcal{M})P(\mathcal M)\\
 & = & \sum_{\mathcal{M}}P(O1\mid pa(O1))P(O5\mid pa(O5))P(O2^{'}\mid pa(O2^{'})P(\mathcal M)\\
\end{eqnarray*}
and
\begin{eqnarray*}
P(\mathcal M) & = & \sum_{\mathcal A} P(\mathcal M\mid \mathcal A)P(\mathcal A)\\
 & = & \sum_{\mathcal A} \prod_{X\in \mathcal M} P(X\mid \mathcal A) \prod_{Y\in \mathcal A} P(Y)\\
 & = & \sum_{\mathcal A}\prod_{X\in \mathcal M} P(X\mid pa(X))\prod_{Y\in \mathcal A} P(Y).
\end{eqnarray*}

Using the assumption that all the association terms of order greater than two were null, 
the likelihood could be computed from all marginal probabilities and  conditional probabilities of order one 
(namely, conditional probability of a variable knowing a unique variable and not a combination of numerous variables). 
For instance, for the node $O2^{'}$, the classical approach required 192 conditional probabilities, whereas our method 
required only seven conditional probabilities.

For the whole BN, in the present application, the number of required probabilities decreased from 381 to 69.
Moreover all required probabilities are now easily interpreted and evaluated by the experts. 
To determine the final probabilities from the evaluations provided by the experts we made use of the rules given in the previous section. 
It can be noticed that by querying more probabilities that necessary allowed us to choose the most reliable probabilities 
in a repeated dialog with the experts.\\

Thus after analyzing the first results on the initial BN, the experts wished to be more precise by adding nine conditional dependences. 
With this new graphical model, the experts suggested to introduce some carefully selected two-order interactions. After they evaluated 
the additional probabilities, the same computation rules were applied to lead to a new Bayesian network. The inference on the resulting BN 
shows that three variables ($Ab$, $Ad$ and $PI3$) appeared to be quite influent on the system degradation.\\

These results encouraged to add maintenance tasks on those three variables in order to improve the reliability of the system. 
The maintenance tasks were easily included in the model as new variables of the BN (see \cite{COR02c}) 
and the experts gave the new conditional probabilities (typically, the probability of an environmental variable knowing a maintenance task) 
by using the methodology previously described. A new inference from this updated BN allowed to give the effect of the 
maintenance actions and to simulate a great number of possible strategies.

\section{Conclusion}

Bayesian Network is a powerful tool to model associations between relevant variables of a problem. 
This kind of modelling requires the intervention of experts. In this work, we concentrated efforts
to provide efficient heuristic methods to get a reliable and meaningful Bayesian network from the 
practical point of view.
First, we applied simple rules to collect information and to design the structure of the graph.
Then, we defined simple and coherent methods to evaluate the probabilities that are needed for inference.
We gave simple rules in order to keep the more reliable probabilities, required to compute the joint probability of the network.
Marginal and conditional probabilities are determined first from operating experience and secondly from expertise. 
One of the main interests of this work which is detailed in \cite{COR03} is to propose a strategy avoiding a too heavy and too unstable acquisition of expert information. This strategy is reducing the number of questions to be asked. Moreover it allows to include maintenance actions as new vertices of the BN (see Corset et al. \cite{COR02c}). Thus, the effect of a maintenance action can be predicted, a point which is rather new and of interest.

\section*{Acknowledgments}
We want to acknowledge the experts of EDF, Fran\c cois Billy, Roger Chevalier, Marie-Agn\`es Garnero, and Jean-Paul Miclot, 
who spent a lot of time to give us the most possible precise information on the system. This work has been achieved when 
G. Celeux and F. Corset were with INRIA Rh\^one-Alpes. We thank the anonymous referee for his comments which helped to improve the presentation.

\bibliographystyle{plain}
\bibliography{BayNet}

\end{document}